\documentclass[conference]{IEEEtran}
\IEEEoverridecommandlockouts
\usepackage{balance}
\usepackage{cite}
\usepackage{amsmath,amssymb,amsfonts}
\usepackage{graphicx}
\usepackage{textcomp}
\usepackage{xcolor}

\usepackage[caption=false]{subfig}
\usepackage{bigfoot}
\usepackage{enumitem}

\DeclareMathOperator*{\argmin}{argmin}

\usepackage{algorithm}
\usepackage{algpseudocode}

\algnewcommand\algorithmicforeach{\textbf{for each}}
\algdef{S}[FOR]{ForEach}[1]{\algorithmicforeach\ #1\ \algorithmicdo}
\algnewcommand{\LineComment}[1]{\State \(\triangleright\) #1}
\algnewcommand\algorithmicswitch{\textbf{switch}}
\algnewcommand\algorithmiccase{\textbf{case}}
\algnewcommand\algorithmicassert{\texttt{save}}
\algnewcommand\Save[1]{\State \algorithmicassert(#1)}%
\algnewcommand\algorithmicsend{\texttt{send}}
\algnewcommand\Send[1]{\State \algorithmicsend(#1)}%
\algdef{SE}[SWITCH]{Switch}{EndSwitch}[1]{\algorithmicswitch\ #1\ \algorithmicdo}{\algorithmicend\ \algorithmicswitch}%
\algdef{SE}[CASE]{Case}{EndCase}[1]{\algorithmiccase\ #1}{\algorithmicend\ \algorithmiccase}%
\algtext*{EndSwitch}%
\algtext*{EndCase}%

\usepackage[hyphens]{url}
\usepackage[T1]{fontenc}
\usepackage[breaklinks,hidelinks]{hyperref}
\newcommand{\reffig}[1]{Fig.~\ref{#1}}

\makeatletter 
\newcommand{\linebreakand}{%
  \end{@IEEEauthorhalign}
  \hfill\mbox{}\par
  \mbox{}\hfill\begin{@IEEEauthorhalign}
}
\makeatother 

\def\BibTeX{{\rm B\kern-.05em{\sc i\kern-.025em b}\kern-.08em
    T\kern-.1667em\lower.7ex\hbox{E}\kern-.125emX}}
\begin{document}

\title{On Kubernetes-aided Federated Database Systems
\thanks{This research is supported in part by startup funding from Queen's University Belfast.}
}

\author{\IEEEauthorblockN{Zheng Li}
\IEEEauthorblockA{\textit{School of EEECS} \\
\textit{Queen's University Belfast}\\
Belfast, UK \\
ORCID: 0000-0002-9704-7651}
\and
\IEEEauthorblockN{Nicol{\'a}s Sald{\'i}as-Vallejos, Mar{\'i}a Andrea Rodr{\'i}guez}
\IEEEauthorblockA{\textit{Department
of Computer Science} \\
\textit{Universidad de Concepci{\'o}n}\\
Concepci{\'o}n, Chile \\
Email: nsaldias2016@udec.cl, andrea@udec.cl}
\and

\IEEEauthorblockN{Austen Rainer}
\IEEEauthorblockA{\textit{School of EEECS} \\
\textit{Queen's University Belfast}\\
Belfast, UK \\
Email: A.Rainer@qub.ac.uk}
}

\maketitle

\begin{abstract}
Cloud computing has made federated database systems (FDBS) significantly more practical to implement than in the past.
As part of a recent Web-based Geographic Information System (WebGIS) project, we are employing cloud-native technologies (from the container ecosystem) to develop a federated database (DB) infrastructure, to help manage and utilise the distributed and various geospatial data. Unfortunately, there seem to be inherent challenges and complexity of applying the container and Kubernetes technologies to building and running DB systems. Considering that most of the geospatial and theme data are pre-obtained and fixed in our WebGIS project, we decided to focus on the read-only user queries and still resort to Kubernetes to implement an FDBS instance to use. Unlike the de facto practices (e.g., using the \texttt{StatefulSets} mechanism, extending Kuberentes APIs, or employing KubeFed), our solution for Kubernetes-aided FDBS simplifies the tech stack by investigating the fractal object of federated data management, inclusively containerising DB instances, and using the lightweight \texttt{Deployment} mechanism to handle stateless DB containers. Overall, this research not only reveals an easy-to-implement approach to constructing read-only components in a fully-fledged FDBS, but also proposes and demonstrates a novel methodology for FDBS investigations. 

\end{abstract}


\section{Introduction}
Database (DB) federation has widely been considered as an ideal solution to the chaos of DB technology sprawl and isolated data islands in large businesses. Unfortunately, due to the various complexities of federated DB systems (FDBS), it was the centralised data warehouse that acted as the superior choice for the vast majority of use cases, until cloud computing emerged as a game changer \cite{Abadi_2021}. By ``disaggregating storage and compute'', the cloud paradigm (especially benefiting from the fast in-cloud network communication) breaks the previously tight coupling between data storing and query processing, which relaxes almost all the challenges associated with query federation and data virtualisation, and thus makes FDBS implementations significantly more practical than in the past.

Driven by a recent Web-based Geographic Information System (WebGIS) project,\footnote{An early proposal project can be found at \url{https://github.com/RodrigoH2Garrido/RodrigoMTV2}} we are also employing cloud-native technologies to implement a federated DB infrastructure, to facilitate the management and utilisation of diverse geospatial datasets. In particular, we try to containerise different DB instances and use Kubernetes to manage the containerised and distributed DB instances, when constructing FDBS blocks. Nevertheless, we met surprisingly many challenges and constraints along this technical path. For example, in addition to the native and tedious configurations (e.g., the Dockerfile and YAML files), the third-party tools may also be required (e.g., the \texttt{Operators} that extend Kubernetes APIs), which often results in a heavy tech stack in practice. There are even dissenting voices against this technical path, such as (i) DB containers should not be used for production \cite{Vsupalov_2022} and (ii) it is not ready yet to run DBs in Kubernetes \cite{Good_2019}.

Based on our successful prototyping work on enabling container-native read-only DB for production \cite{Li_2021}, we decided to keep emphasising read-only queries and still implement a Kubernetes-aided FDBS to use. From the application's perspective, most of the  geospatial and theme data (e.g., postal code) are pre-obtained and fixed in our WebGIS project. In fact, more than half of the data used by real-world applications are read-only \cite{Vaumourin_2014}. From the technological perspective, our container-native read-only DB solution significantly simplifies the tech stack for building a Kubernetes-aided FDBS. On the one hand, the container-native feature strictly guarantees the shared-nothing architecture of an FDBS, and then enjoys the architectural advantages (e.g., the increased scalability and the reduced blast radius of failures). On the other hand, the read-only feature naturally avoids the complexity and overhead of handling stateful services in Kubernetes.

By identifying the fractal pattern of, and by analysing the fractal object of data management in FDBS, this paper explains our system design together with  demos and prototypes of the relevant technical implementations. Those demos and prototypes in turn validate the system design.\footnote{We share the documentation about a demo component system of our proposed Kubernetes-aided FDBS at \url{https://zenodo.org/record/6991105}.} Correspondingly, this research makes a twofold contribution:

\begin{itemize}
\item For practitioners, our work offers a handy tutorial for implementing Kubernetes-aided FDBS. Although this solution is only applicable for read-only user queries, it can be a lightweight and solid cornerstone to support read-only components in a fully-fledged FDBS.
\item For researchers, our work reveals a novel methodology for FDBS investigations, i.e.~studying the fractal pattern and the fractal object of FDBS. This approach further opens new research opportunities, e.g., calculating the fractal dimension to measure complexity and perform comparison between different FDBS instances. 
\end{itemize}

The remainder of this paper is organised as follows. Section \ref{sec:relatedwork} briefly introduces the background of FDBS and particularly highlights the current practices of applying Kubernetes to building federated systems. Section \ref{sec:motivation} describes the motivation project of this research. By focusing on the identified fractal object, Section \ref{sec:design} specifies the design and initial implementation of our Kubernetes-aided FDBS. Finally, we discuss the threat to validity, draw conclusions and give our future work plans in Section \ref{sec:conclusions}.

\section{Background and Related Work}
\label{sec:relatedwork}
As a special type of multi-DB systems, FDBS are characterised by the hierarchical management of autonomous component DB systems \cite{Sheth_1990}. The hierarchical and autonomous DB management has become more and more valuable and necessary for modern applications, because of the increasingly involved heterogeneous datasets with diverse models, ownership, and storage technologies \cite{Azevedo_2020}. A typical example is the GIS domain where various geospatial data may be acquired and utilised by different organisations via different acquisition schemes and channels \cite{Butenuth_2007}. Thus, partial and controlled data sharing across those organisations would be more practical than building up a centralised data lake. An advantage of GIS applications in this case is that the standard geographical coordinates can naturally act as universal foreign keys to facilitate the geospatial data sharing and integration. 
This also motives our research described in this paper (see Section \ref{sec:motivation}).

\subsection{Kubernetes Federation}
Kubernetes is the de facto open-source platform for automatically deploying, scaling, and managing containerised applications. Benefiting from the convenient and automated Kubernetes mechanism of arranging applications within a single cluster, Kubernetes Cluster Federation (a.k.a.~KubeFed) was developed to coordinate and manage multiple clusters. Although still at a developing stage, KubeFed has been employed to study pioneer solutions for distributing application components and for balancing workloads in the modern computing paradigms \cite{Faticanti_2021}. Meanwhile, more research interests can be found in improving KubeFed. For example, a policy-based scheduling architecture is proposed as a supplement to KubeFed for determining the appropriate federated clusters \cite{Kim_Kim_2021}; whereas some other researchers argue to replace the KubeFed control plane with a decentralised control plane for the federation of Kubernetes clusters \cite{Larsson_2020}.   

In contrast with KubeFed, we claim two significant differences in this research. Firstly, KubeFed is a top-down strategy to orchestrate the federated infrastructure composed of known clusters, while our research is a bottom-up approach to growing a federated (DB) system without necessarily knowing its eventual size. When using KubeFed, a configuration template needs to be generated in a host cluster and will be propagated to the pre-determined member clusters. As mentioned in \cite{Larsson_2020}, this strategy weakens the autonomy of the federation components and inevitably leads to a single point of failure. In our research, each component system is supposed to autonomously manage its own infrastructure and configuration (e.g., load-balancing policies), and to stay unaware of its involvement in any upper federated system(s). 


Secondly and more importantly, KubeFed targets the application federation, while our work focuses on the DB federation. Considering that the application containerisation in the production environments does not containerise DBs (instead, fully managed cloud DB services are used \cite{Heddings2_2020}), the Kubernetes-based application federation does not guarantee the existence of an underlying FDBS. Consequently, scaling an application within the federated Kubernetes clusters could cause performance bottlenecks or require extra/external DB service adjustments, due to the mismatch between the data accessing points (at the DB level) and the data processing points (in the containerised modules).  

Given these differences, we further argue that this research and KubeFed would reinforce each other and eventually be able to be combined together. Following the layered software architecture, KubeFed's top-down strategy can still be suitable for globally optimising the deployment of given applications, while our bottom-up approach may be particularly used for integrating the diverse and unpredictable datasets.

\subsection{Running Databases in Kubernetes}

As mentioned above, when containerising and managing an application that involves DBs, the common practice in production is to host and maintain the DBs independently outside of Kubernetes \cite{Larsson_2019,Perera_2021}. Google engineers confirm that it is still far from running a DB in Kubernetes in a practical sense, and we have to wait for the further evolution of relevant technologies and tools  \cite{Good_2019}. 

Despite those majority opinions, the growing container ecosystem keeps fostering interests in running DBs in Kubernetes, and we have observed two study paths: 

\begin{itemize}
\item The first path advocates the \texttt{StatefulSets} mechanism of Kubernetes to enable the state management for DBs, by automatically provisioning persistent volumes on a per pod basis \cite{Delnat_2018}. However, using external volumes for data persistence could encounter dangling volume issues \cite{Heddings2_2020} and will introduce disk I/O performance bottlenecks \cite{Truyen_2019}. 
\item To address the limits of \texttt{StatefulSets}, the second path employs third-party \texttt{Operators} to extend Kubernetes APIs, by creating custom resources and controllers \cite{Perera_2021}. However, the current \texttt{Operators} are immature and unsuitable for mission-critical workloads \cite{Larsson_2019}, not to mention that the extended tech stack may even worsen the complications of running DBs in Kubernetes. 
\end{itemize}

Different from both of the study paths, our research argues to simplify the tech stack, by employing a container-native data persistence solution, and then using the straightforward \texttt{Deployment} mechanism to run read-only DBs in Kubernetes. The simplified tech stack not only enables Kubernetes to inclusively handle read-only DB clusters, but also encourages us to work further on Kubernete-aided FDBS in production, at least for stateless applications at this current stage. To our best knowledge, this is the first study that employs Kubernetes to investigate and implement FDBS.



\section{The Motivation Project}
\label{sec:motivation}
The motivation of this research is a recent WebGIS project of ours. This WebGIS project aims to build up an efficient DB infrastructure, to facilitate offering online GIS services. 
By enabling collecting, storing, managing, analyzing, and displaying geospatial data, GIS has become an indispensable software tool for studying the scientific discipline Geography \cite{Han_2019}. Benefiting from the significant development of the Internet, modern GIS implementations are increasingly embracing the web technologies to provide low-cost and platform-independent services \cite{Li_GIS_2020}. Then, users can conveniently consume the online services that are comparable to those of desktop GIS \cite{Fast_2020}, without necessarily installing any GIS software or maintaining the geographical data.

However, the service response speed has been identified as a performance bottleneck that may harm the user experience of WebGIS. In addition to the possibly heavy communication overhead over the Internet \cite{Sahin_2021}, the major problem is essentially the monolithic software infrastructure, which ``leads to the inefficiency of GIS software in the face of data intensity and computation complexity'' \cite{Zhu_Zhao_2021}.
It should be noted that the widely discussed client/browser-server architecture and the multi-layer system design do not guarantee a monolith-free implementation of WebGIS. For example, the bottom layer will still be monolithic if a single spatial DB is used to support the whole system \cite{Vaitis_2019}. 
 In practice, multiple simultaneous clients may request different parts of a large geographic dataset to be rendered with different analysis functions \cite{Meyer_2007}.

Therefore, we advocate the federated architecture for breaking the monolith of WebGIS implementations. In particular, many geospatial and theme data (e.g., postal code) we plan to use are fixed in advance, not to mention the historical data that are not allowed to change in general. For the historical data, different periods of time series may also be stored and processed separately, to facilitate the data management and visualisation. For example, the PRISM Climate Group\footnote{PRISM Climate Data: \url{https://prism.oregonstate.edu/}} distinguishes between Recent Years and Historical Past when offering temperature and precipitation data at multiple spatial/temporal resolutions \cite{Daly_2015}. Thus,
we naturally recall the read-only DB container solution \cite{Li_2021} to develop a Kubernetes-aided FDBS for our WebGIS project.

\section{The Design and Initial Implementation of Kubernetes-aided Federated Database Systems Based on a Fractal Pattern}
\label{sec:design}
\begin{figure}[!t]
\centerline{\includegraphics[trim=58 668 370 52,clip]{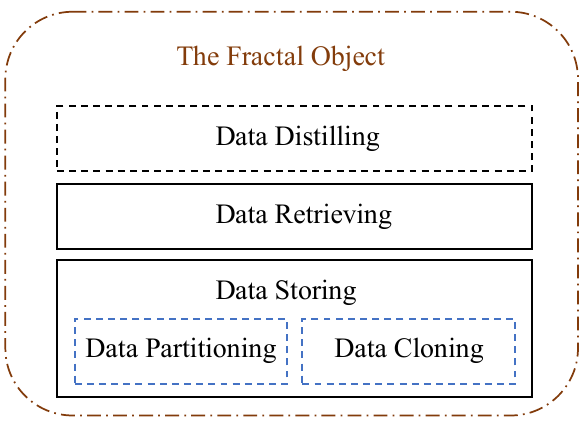}}
\caption{The fractal object of data management in federated database systems.}
\label{fig:object}
\end{figure}

Given the definition of FDBS, we identify a fractal pattern from the data management's perspective: An FDBS is composed of a set of autonomous components, and each component consists of multiple autonomous sub-components also in the federated way, which repeats until all the leaf components are single DB instances. By emphasising the basic data management activities, we clarify the self similarity of the fractal to be the combination of data storing, data retrieving, and (optional) data distilling. Particularly, data storing may also come with data partitioning and/or data cloning. As such, the fractal object of data management can represent either a constituent DB, a component DB system, or an FDBS at any scale, as illustrated in \reffig{fig:object}.

Using our motivation project as the context, we elaborate this fractal object by explaining the involved data management activities in the following subsections.

\subsection{Data Distilling}
To maximise the alignment with the literature, we use the term ``data distilling'' instead of ``data analytics'' to indicate all kinds of operations/methods that make sense of the retrieved data, ranging from redundancy reduction \cite{Rios_Lin_2012} to information/knowledge extraction \cite{Frankel_2008}. From a single DB's perspective, data distilling can be distinguished from the back-end DB system and merged into upper-layer business logic implementations. When it comes to concurrently distributed DBs, the DB integration technologies also emphasise data usage (distilling) at an abstract level, via standard user interfaces and uniform data access \cite{Ozsu_2020,Stojanovic_2022}. 
However, data distilling is not separable from the fractal object of FDBS, even though it can be optionally implemented or ignored by different component systems. Driven by the distinctive feature (namely autonomy) of federated architecture, the component systems of an FDBS should be completely self-sustained and functional, i.e.~a component can not only respond to the upstream retrieval requests, but can also support its own self-sufficient data distilling tasks. 

Our WebGIS project follows the distillation use case of \cite{Rios_Lin_2012} to synthesise and visualise the geospatial data. A typical functionality is to group and/or identify neighbours of the United States' postcodes on a map at different zoom levels. Ideally, the United States' postcode visualisation system can be supported by distributed DBs belonging to the individual states, while each state may customise a standalone visualisation system based on its own data for the local usage. Meanwhile, the distributed DB system of the United States can also contribute the national data to a world-wide postcode system for the global usage.

\begin{figure}[!t]
\centerline{\includegraphics[width=0.9\columnwidth,height=4.5cm]{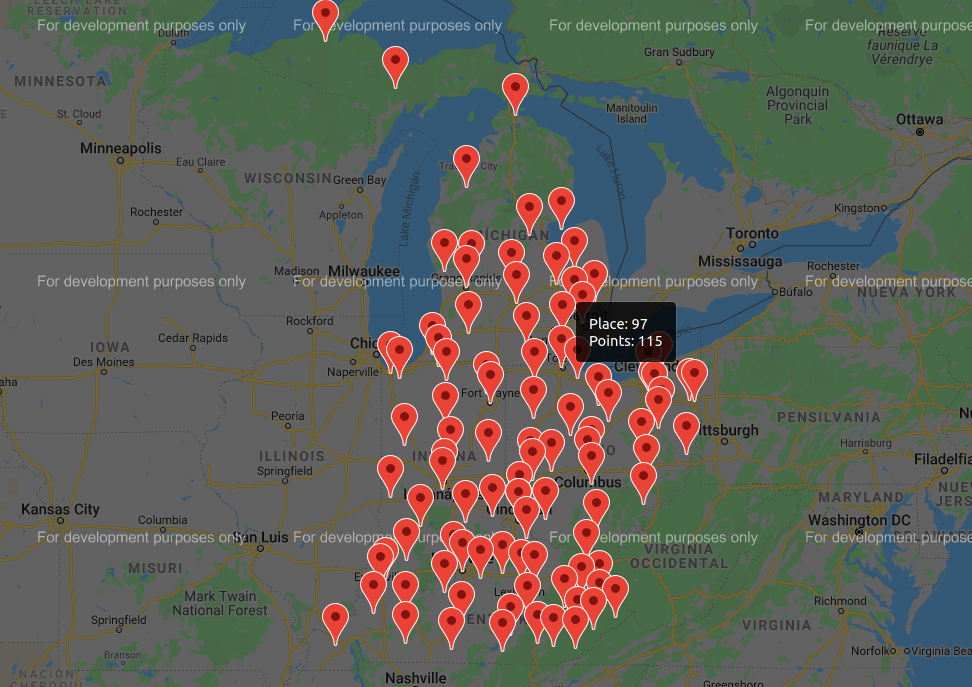}}
\caption{The distribution of the United State's postcodes starting from ``4''.}
\label{fig:map}
\end{figure}

In practice, for the convenience of implementation and demonstration, we take advantage of the characteristics of postcodes and divide them into ten DBs according to their first digits (i.e.~zero, one, ..., nine). The grouping rules and the neighbour identification strategy are implemented into a serverless function with Amazon Web Services (AWS) Lambda,\footnote{At the time of writing, we are migrating the AWS Lambda function to a microservice based on Amazon Elastic Kubernetes Service (Amazon EKS), in order to take advantage of the  catalog of Kubernetes services of distributed DBs (see Section \ref{subsec:retrieving}).} as shown in Algorithm \ref{postcode_algorithm}.  The user interface source files (e.g., the HTML pages, self-defined marker images, etc.) are deployed to the Amazon Simple Storage Service (Amazon S3). It should be noted that the Lambda function and the interface are independent from the back-end DBs, and thus they can be used both for the whole DB system and for any component DB. For example, \reffig{fig:map} demonstrates the usage of a single DB to support visualising the distribution of the United State's postcodes starting from ``4'' only.

\begin{algorithm}[!t]\footnotesize
  \hspace*{\algorithmicindent} \textbf{Input:} Zoom level \textit{zoom}, the optional inputs \textit{address} $= (0,0)$ and its neighbors' number $k=-1$.\\
  \hspace*{\algorithmicindent} \textbf{Output:} A set of geographic locations (i.e.~centroids or the closest $k$ ones to the \textit{address}) to be visualised.
  \caption{Identifying Geographic Locations}\label{postcode_algorithm}
\setlength\baselineskip{9pt}
  \begin{algorithmic}[1]  
  \LineComment{Grouping postcodes based on zoom levels on the map.}
  \State \textit{Groups} $\gets \emptyset$
  \Switch{\textit{zoom}} 
  
  \setlength\baselineskip{1em}
      \Case{1, 2, 3, 4:} 
           \State \textit{Groups} $\gets$ \textbf{group} postcodes \textbf{by} the \textit{first} digit.
      \EndCase
      \Case{5, 6:} 
          \State \textit{Groups} $\gets$ \textbf{group} postcodes \textbf{by} the \textit{first two} digits.
      \EndCase
      \Case{7, 8:}
           \State \textit{Groups} $\gets$ \textbf{group} postcodes \textbf{by} the \textit{first three} digits.
      \EndCase
      \Case{9, 10:}
           \State \textit{Groups} $\gets$ \textbf{group} postcodes \textbf{by} the \textit{first four} digits.
      \EndCase
      \Case{others:}
           \State \textit{Groups} $\gets$ \textbf{group} postcodes \textbf{by} the \textit{all five} digits. \Comment{No grouping.}
      \EndCase
  \EndSwitch 
  
  \setlength\baselineskip{10pt}
  \State \textit{Centroids} $\gets \emptyset$
  \State \textit{Centroids} $\gets \{\text{Centroid coordinates of~}\textit{group} \mid \textit{group}\in \textit{Groups}\}$

   \If{$k<0$} 
        \LineComment If \textit{address} and $k$ are not specified, the functionality is grouping.
        \State \Return \textit{Centroids}
   \EndIf

  \LineComment{Calculate distances between \textit{address} and the centroids.}
  \State \textit{Distances}$ \gets \emptyset$
  \State \textit{Distances} $\gets \{(\textit{centroid}, |\textit{centroid}-\textit{address}|) \mid \textit{centroid}\in \textit{Centroids}\}$
  
  \LineComment{Select $k$ centroids that are the $k$ nearest neighbours of \textit{address}.}
 \State \textit{Neighbours}$ \gets \emptyset$
  \State \textit{Neighbours} $\gets \argmin_{D\subset \textit{Distances}, |D|=k}\sum_{c\in D.\textit{centroids}}D[c]$ 
  
   \setlength\baselineskip{12pt}
      \State \Return \textit{Neighbours.centroids} \Comment{$\{c \mid (c, \textit{Distances}[c])\in \textit{Neighbours}\}$}
  \end{algorithmic}
\end{algorithm}
 
\subsection{Data Retrieving}
\label{subsec:retrieving}
Data retrieving and storing correspond to a pair of opposite basic DB operations, i.e.~reading and writing data. In generic scenarios, these two operations coexist when satisfying user requests. In our WebGIS project, we decouple these two operations by constraining user requests to be data retrieving only, while leaving data storing to the role of DB administrators (see~Section \ref{subsec:storing}). This is both to reflect the characteristics of our current GIS datasets and to match the features of FDBS. As mentioned previously, many GIS-related data barely change and are maintained by different stakeholders rather than by the end users. For example, the postcodes in a country must be allocated by the government or by the government-authorised entities, and once allocated they can only be used but not changed by the residents. The requirement of managing such datasets further justifies the suitability of FDBS for our WebGIS project, as an FDBS accepts different ownership of its individual components instead of using centralised control \cite{Kim_Moon_2018}.

\begin{figure}[!t]
\centerline{\includegraphics[trim=60 557 315 60,clip]{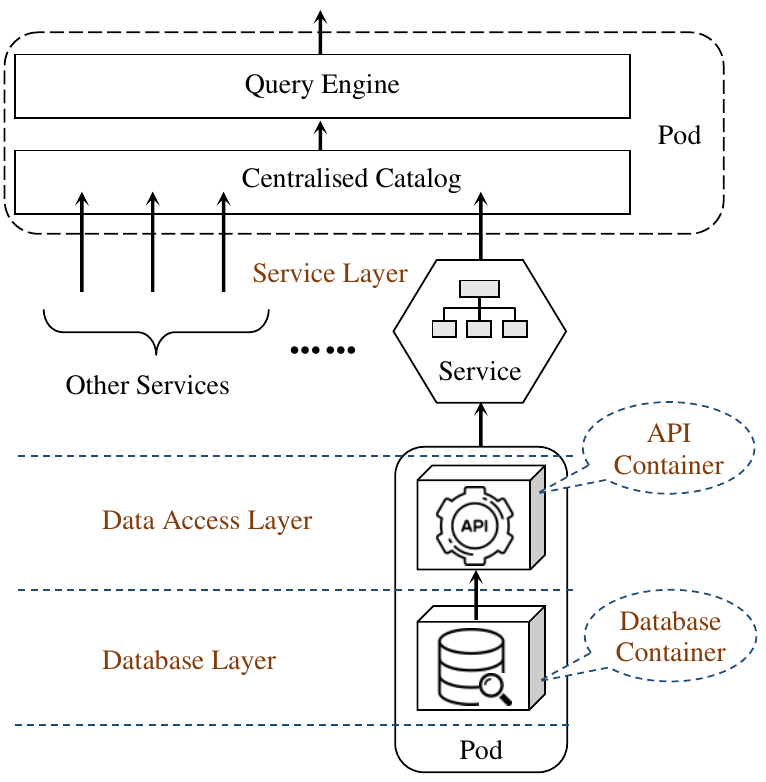}}
\caption{The essential parts and architecture for data retrieving implementation.}
\label{fig:retrieving}
\end{figure}

When it comes to employing Kubernetes technologies to implement data retrieving, we highlight the essential parts and our implementation architecture by demonstrating a single DB scenario, as portrayed in \reffig{fig:retrieving} and explained below.

\begin{itemize}
\item \textbf{Self-contained DB container}. Considering that there is no data input from the end users, we have developed an inclusively containerised read-only DB solution (see more details in Section \ref{subsec:storing}). Although we only used MySQL for the demonstration \cite{Li_2021}, the containerisation is not limited to any DB product, not to mention the publicly accessible cloud DB services. In other words, our solution is compatible with the design autonomy of FDBS that allows the component systems to freely choose DB technologies, query languages, and data models \cite{Doncevic_2020}.

\item \textbf{Data access API container}. Given the heterogeneity in DBs (that is also the nature of FDBS), we supplement a containerised data access layer and use RESTful APIs to hide those connection routines and query mechanisms. Note that this data access layer does not violate the aforementioned design autonomy, because RESTful APIs can also be realised using flexible languages and frameworks (e.g., Node.js and Flask). On the contrary, this data access layer facilitates realising the execution autonomy of constituent DBs, by exposing suitable APIs to control the operations from external requests. 

\item \textbf{Pod}. A pod is the smallest deployable and manageable unit in Kubernetes. We define the DB container and the corresponding API container to be co-located in a single pod. Thus, the tight coupling guarantees any deployed DB and its data access layer to either work along or be gone together. Pods have to run on one or multiple nodes that may be virtual or physical machines. For the purpose of conciseness, we intentionally omit the environmental element (where the Kubernetes concept \texttt{Node} exists) from the illustration (see~\reffig{fig:retrieving}).

\item \textbf{Service}. Despite the data access layer that has provided APIs, we still need a Kubernetes service layer for data retrieval, for two reasons:

\begin{enumerate}[topsep=0.2em,leftmargin=2\parindent]
\item The pods have an ephemeral nature as well as the containers at runtime. Although Kubernetes can recreate new ones if any pods are lost, 
the addresses of the pods (or the hosts of APIs) will unavoidably change. Therefore, an abstract service with a fixed IP address will be needed to reconcile the uncertainty, by automatically selecting the pods (including newly created ones) with predefined labels. 
\item Kubernetes services enable a unified connection mechanism between an FDBS and its components, and also between the components. A service can not only abstract the definition and the access policy of pods, but can also abstract other kinds of back-end endpoints.\footnote{\url{https://kubernetes.io/docs/concepts/services-networking/service/#services-without-selectors}} For example, given a standalone query engine (see \textbf{Query engine} below) that offer single query capabilities for a federated component system, we can manually create an \texttt{Endpoints} object for it and map its network address to a standard Kubernetes service. Note that the behaviours of such an engine service will have no difference to the basic DB services.
\end{enumerate}

\item \textbf{Centralised catalog of services}. Once an aforementioned service is exposed, it is ready to be used to retrieve data from its back-end DB(s). In other words, the service can be viewed as an equivalent representation of its back-end DB(s). Recall the definition that an FDBS is a heterogeneous collection of DBs \cite{Azevedo_2020}. In this case, the FDBS may also be considered as a service collection. To drive multiple and distributed services to co-work as an autonomous entity, a prerequisite is to make them conveniently discover-able, and we advocate to use a centralised catalog of services. The catalog includes not only the service addresses and callable APIs, but also the respective metadata about the data sources that can facilitate the service usage. 

\item \textbf{Query engine}. Based on an available and centralised catalog of DB services, we need a query engine to interpret, optimise, and break external requests into internally serviceable queries. Since the underlying data access is based on APIs and the data models have been reflected (or constrained) by the API paths and parameters, we summarise the knowledge of the service catalog and further design abstract APIs to act as the query engine, instead of resorting to a traditional SQL engine (that is also inappropriate here). Furthermore and optionally, by sticking to the fractal point of view, the centralised catalog and the query engine can be similarly implemented as a DB and its access layer. Then, it will be convenient to follow the same routine to respectively containerise them, tightly couple them into a pod, and expose the pod via a Kubernetes service.
\end{itemize}

\subsection{Data Storing}
\label{subsec:storing}
Before being able to retrieve data, there must be data already stored persistently in accessible repositories. 
Similar to Section \ref{subsec:retrieving}, we start explaining our data storing mechanism also from a single DB's perspective. Since our project only focuses on the reading requests from users, we have employed a container-native read-only DB solution \cite{Li_2021} for storing the spatial datasets we need. Unlike the conventional DB containerisation solution that relies on external volumes for data persistence, this container-native solution pre-bakes the necessary data into DB images. Given the immutable nature of image, the pre-baked data will be unchangeable for use at runtime. When the involved datasets need to be updated, the administrator who is in charge of the data source can rebuild DB images to replace the old ones. In practice, a local DB may be maintained to generate DB dumps to facilitate the image building.
Based on new images, when data cloning is implemented (see~Section \ref{subsubsec:cloning}), the rolling update techniques \cite{Rossi_2020} can further be employed to update a cluster of DB containers smoothly and directly in the production environment.


In addition to satisfying the read-only needs, the container-native approach to data persistence can bring various benefits, ranging from avoiding the issue of dangling volumes to improving the performance of data retrieval. 
However, this DB container solution is not encouraged to cater large-size datasets, because the oversized images will significantly reduce the portability and increase the deployment latency of the corresponding containers.
Therefore, a practical strategy of dealing with a large dataset is to break the dataset into small subsets and store them via multiple DB containers. It is noteworthy that this strategy well matches a foundational idea of FDBS,  i.e.~enabling the divide-and-conquer management and organisation of data.

\begin{figure}
     \centering
     \subfloat[The data partitioning scenario.
        \label{subfig:partitioning}]{\includegraphics[trim=65 670 337 65,clip]{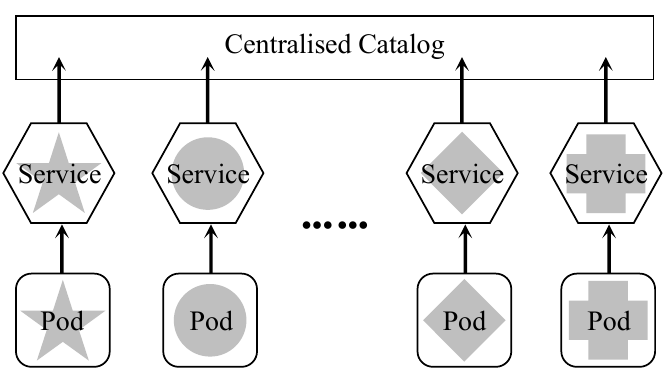}}
     \vfill
     \subfloat[The data cloning scenario.
        \label{subfig:cloning}]{\includegraphics[trim=65 670 337 65,clip]{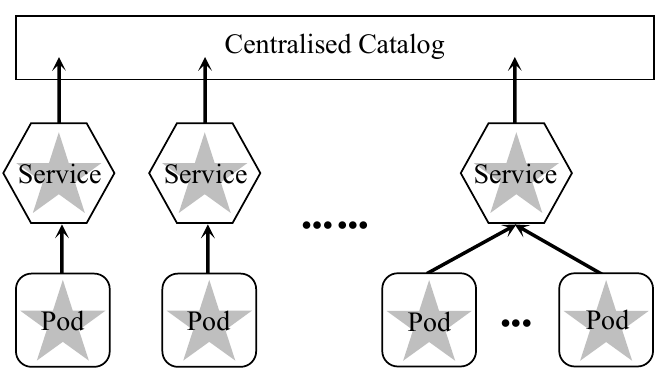}}
     \vfill
     \subfloat[The mixed scenario including both data partitioning and cloning.
        \label{subfig:mixed}]{\includegraphics[trim=50 670 322 65,clip]{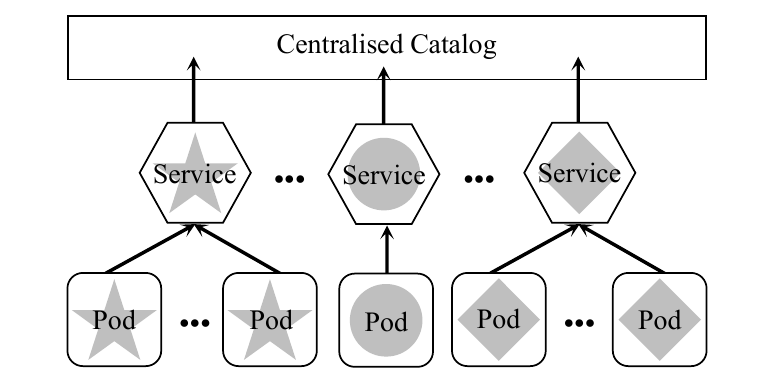}}
        \caption{Three multi-database scenarios for data storing.}
        \label{fig:scenarios}
\end{figure}

Correspondingly, after fixing the mechanism for storing data with respect to a single DB, we move on to the discussion about data storing in the multi-DB scenarios. In detail, we distinguish between three scenarios, namely data partitioning, data cloning, and a mixed scenario that involves the both, as abstractly illustrated in \reffig{fig:scenarios}. Particularly, since data cloning can be implemented at both the \textit{pod} level and the \textit{service} level in Kubernetes (see Section \ref{subsubsec:cloning}), we highlight only these two Kubernetes parts in the illustration, and using different (or the same) geometric shapes to respectively represent different data (or the identical clones) that are stored in DB containers and are retrievable via services.

\subsubsection{The Data Partitioning Scenario}
Despite having different motives, data partitioning is the inherent requirement of both FDBS and our container-native DB solution. Here we consider the originally distributed datasets as the result from data partitioning by nature, while the distributed datasets may need further splits for non-functional purposes. For example, the Open Geospatial Consortium (OGC) standards\footnote{OGC Standards: \url{https://www.ogc.org/docs/is}} define the most fundamental GIS modules as \textit{simple features} who provide a common way to store and access geographical feature data. Thus, it is natural to associate a whole feature dataset with a \textit{simple feature} implementation. When the \textit{simple feature} involves a large amount of data, we advocate to divide the whole dataset into suitable subsets to balance, and to take advantage of, the DB containers' portability (see~\reffig{subfig:partitioning}). 

\begin{figure}[!t]
\centerline{\includegraphics[trim=55 580 350 65,clip]{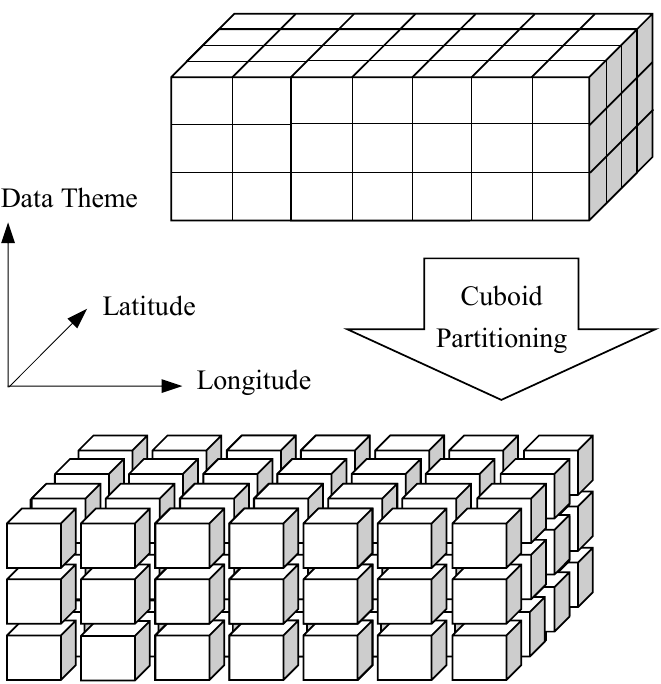}}
\caption{Partitioning spatial data according both to their geographic coordinates and to their themes.}
\label{fig:cuboid}
\end{figure}

We have proposed a cuboid partitioning approach to help identify suitable sub datasets, as illustrated in \reffig{fig:cuboid}. On the one hand, given the two-dimensional representation of the Earth's surface, it is convenient and flexible to slice the Earth map into different sizes of pieces, especially for the situation when the geographical data are sparse or unevenly distributed. On the other hand, we regulate that different themes of data must be separated and stored in different DBs. The data themes can be either from nature (e.g., climate, forest biomass, underground resources, etc.) or from the human world (e.g., postcode, power grid, traffic flow, etc.). The effects of this cuboid partitioning approach has already been reflected by the aforementioned demo of data distilling (see~\reffig{fig:map}).

\subsubsection{The Data Cloning Scenario}
\label{subsubsec:cloning}
To address the throughput bottleneck of DB or network (e.g., due to request floods), increasing data/DB redundancy is a common tactic to improve the performance and reliability of DB systems. Thus, it is worth considering to create and separately store data clones after the data partitioning is done. When employing the Kubernetes technologies to conduct data cloning and manage the data clones, we distinguish between the implementation of parallel DB services and the implementation of a DB pod cluster, as represented respectively by the left two services and by the rightmost service in \reffig{subfig:cloning}. 

As the name suggests, a DB pod cluster clones the stored data by replicating the pods. Considering that read-only DBs are carefree about the state information, our two-container pod (see Section \ref{subsec:retrieving}) can be viewed as hosting a stateless application. In the stateless situation, Kubernetes can use a declarative \texttt{Deployment} object to create and automatically manage a set of pod replicas. 
To avoid repeating the official documentation, we do not re-explain the technical details of Kubernetes \texttt{Deployment} mechanism here.  

\begin{figure}[!t]
\centerline{\includegraphics[trim=5 5 5 5,clip]{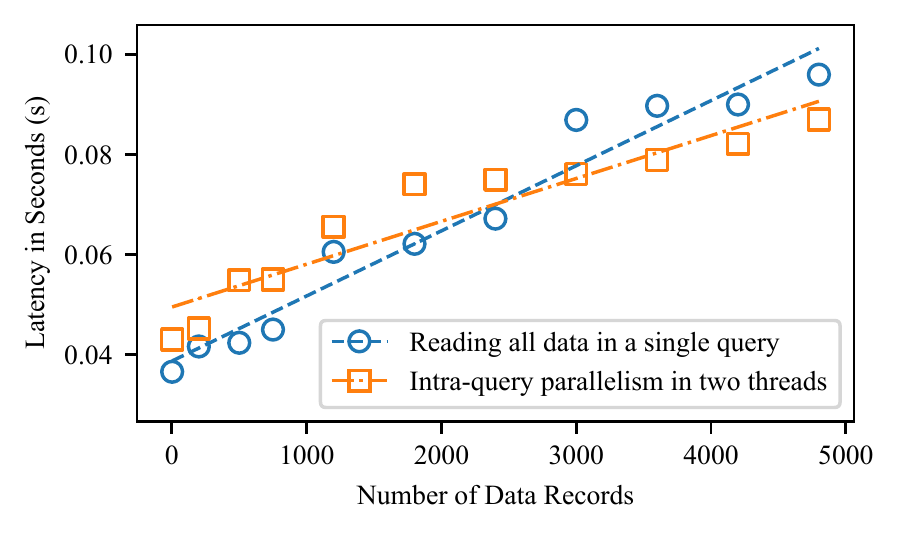}}
\caption{Linear regression modeling analysis and performance comparison of the intra-query parallelism in two threads against single-query data reading.}
\label{fig:model}
\end{figure}

Since the abstraction of a DB pod cluster is still a single DB service, the parallel DB services can in turn be viewed as a set of DB pod clusters, which is essentially a special instance of the multi-cluster deployment architecture \cite{Wickramasinghe_2021}. Compared with a single cluster of pods handled by Kubernetes, the multi-cluster deployment would need extra and inter-cluster load-balancing techniques. This, however, gives us more flexibility and controllability to improve the performance of a multi-cluster system. For example, we have investigated the effectiveness and efficiency of intra-query parallelism against the cloned data, i.e.~decomposing a single query into smaller tasks and executing them concurrently \cite{Hardavellas_2009}. By using the benchmark dataset of Amazon's spot service price history,\footnote{\url{https://doi.org/10.5281/zenodo.4583507}} we demonstrate the performance measurement and comparison between the single-query data reading and the two-thread intra-query parallelism, as shown in \reffig{fig:model}. As can be seen in this case, it would be worth breaking a single query evenly into two sub-queries, when the data to be retrieved have roughly more than 3000 records. Note that our measurement has included the time consumed for pre-counting the needed data records (by using a pre-query with the SQL statement starting from \texttt{SELECT COUNT()}). 

This demo proves that we can take advantage of the pre-knowledge of both the cloned data and their storage architecture (e.g., empirically building an intra-query parallelism lookup table with respect to the concurrency level and workload size), to actively improve the data retrieval performance, rather than passively rely on the built-in load balancing of Kubernetes. Ideally, the evaluation and optimisation of a sizeable read-only query can be implemented similar to the peer-to-peer data sharing \cite{Agarwal_2012} (from a single node's perspective), i.e.~concurrently reading different data segments from many peer and identical DB instances in an FDBS.

\subsubsection{The Mixed Scenario Including both Data Partitioning and Cloning}
In practice, we would need to have data partitioning and cloning mixed together for a real-world FDBS and its standalone component systems, to support multi-feature applications. Thus, this scenario represents the generic situation of, and the two-dimensional space of data storage scaling. For example, a component system may involve multiple Kubernetes services for different DB instances, while each service umbrellas a set of Kubernetes pods that are clones of a specific DB instance, as illustrated in \reffig{subfig:mixed}. Correspondingly, this scenario also indicates the needs and opportunities of a global optimisation, e.g., taking into account the DB containers' portability, the data availability/reliability, and the overall query performance, as explained in the previous scenarios.




\section{Discussion and Conclusions}
\label{sec:conclusions}
\subsection{Threat to Validity}
The major threat to validity of our FDBS solution is its application constraint, i.e.~it does not allow the end users to input or modify any data. Therefore, we remind readers to do an applicability analysis before employing this FDBS solution. 

On the other hand, we argue that the constraint has been naturally relaxed to some extent, for the existence of read-only data everywhere in our normal life, e.g., the catalog of product types and the customers' shopping history in a supermarket \cite{Ghilardi_2022}. Some applications can completely be based on the read-only datasets, e.g., the online dictionary\footnote{\url{https://dictionary.cambridge.org/}} and the postal code lookup wizard\footnote{\url{https://website-uat.correos.cl/codigo-postal}}. More specifically, by analysing the standard benchmarks for embedded systems, it is estimated that up to 62\% of all the data used by applications are read-only \cite{Vaumourin_2014}.

Moreover, we can expect further benefits by setting a database to the read-only mode for read-only data. From the perspective of database administrators, the overhead and issues concerning transactions and concurrency control such as locking \cite{Kyte_2014} and multi-version \cite{Faleiro_2015} will be avoided, as no locks are needed for any query from a read-only database. From the perspective of programmers, the multi-process data access will be simplified, without necessarily concerning about message passing or shared memory programming \cite{Butler_2015}. 

\subsection{Conclusions and Future Work}
As part of a recent WebGIS project, we are designing and developing a federated DB infrastructure to facilitate managing and utilising various and distributed geospatial data.
Given our successful prototyping work, we are confident to conclude that it is feasible and practical to employ Kubernetes in FDBS implementations for read-only user queries. It should be noted that our conclusion does not deny the consensus about the current technical limits of Kubernetes \cite{Good_2019}. We acknowledge that our solution essentially bypasses, rather than fixes, the challenges of running DBs in Kubernetes and the complexity of Kubernetes-based federation. However, this lightweight solution can efficiently satisfy the needs of read-only components in a fully-fledged FDBS, no matter how the Kubernetes technologies will evolve.    

Benefiting from our current research outcomes, we are able to unfold the future work along two directions. First of all, we will gradually finalise the Kubernetes-aided FDBS infrastructure along with the development of our WebGIS project. Then, based on our first-hand experience, we plan to undertake more formal analyses of the fractal pattern and the fractal object of FDBS, aiming to develop new theories and/or guidelines for FDBS implementations.

\bibliographystyle{IEEEtran}
\bibliography{FDBRefShort}

\end{document}